\documentclass[twocolumn,preprintnumbers,amsmath,superscriptaddress,amssymb,aps]{revtex4-1}

\usepackage{graphicx}
\usepackage{dcolumn}
\usepackage{bm}
\usepackage{amsmath}
\usepackage{amssymb}
\usepackage{graphicx}
\usepackage{indentfirst}
\usepackage{booktabs}
\usepackage{multirow}
\usepackage{colortbl}

\def\degree{${}^{\circ}$}
\selectfont

\begin{document}

\title{Large magneto-optical effects and magnetic anisotropy energy in two-dimensional Cr$_2$Ge$_2$Te$_6$}

\author{Yimei Fang}
\address{Department of Physics, Collaborative Innovation Center for Optoelectronic Semiconductors 
and Efficient Devices, Key Laboratory of Low Dimensional Condensed Matter Physics 
(Department of Education of Fujian Province), Jiujiang Research Institute, Xiamen University, Xiamen 361005, China}
\author{Shunqing Wu}
\email{wsq@xmu.edu.cn}
\address{Department of Physics, Collaborative Innovation Center for Optoelectronic Semiconductors 
and Efficient Devices, Key Laboratory of Low Dimensional Condensed Matter Physics 
(Department of Education of Fujian Province), Jiujiang Research Institute, Xiamen University, Xiamen 361005, China}
\author{Zi-Zhong Zhu}
\email{zzhu@xmu.edu.cn}
\address{Department of Physics, Collaborative Innovation Center for Optoelectronic Semiconductors 
and Efficient Devices, Key Laboratory of Low Dimensional Condensed Matter Physics 
(Department of Education of Fujian Province), Jiujiang Research Institute, Xiamen University, Xiamen 361005, China}
\author{Guang-Yu Guo}
\email{gyguo@phys.ntu.edu.tw}
\address{Department of Physics and Center for Theoretical Physics, National Taiwan University, Taipei 10617, Taiwan}
\address{Physics Division, National Center for Theoretical Sciences, Hsinchu 30013, Taiwan}


\date{\today}

\begin{abstract}
Atomically thin ferromagnetic (FM) films were recently prepared by mechanical exfoliation
of bulk FM semiconductor Cr$_2$Ge$_2$Te$_6$.  
They provide a platform to explore novel two-dimensional (2D) magnetic phenomena,
and offer exciting prospects for new technologies.
By performing systematic {\it ab initio} density functional calculations, 
here we study two relativity-induced properties of these 2D materials [monolayer (ML), 
bilayer (BL) and trilayer (TL) as well as bulk], namely, magnetic anisotropy energy (MAE) 
and magneto-optical (MO) effects.
Competing contributions of both magneto-crystalline anisotropy energy (C-MAE) and magnetic 
dipolar anisotropy energy (D-MAE) to the MAE, are computed. Calculated MAEs of
these materials are large, being in the order of $\sim$0.1 meV/Cr. 
Interestingly, we find that the out-of-plane magnetic anisotropy 
is preferred in all the systems except the ML where an in-plane magnetization
is favored because here the D-MAE is larger than the C-MAE. Crucially, this explains why long-range
FM order was observed in all the few-layer Cr$_2$Ge$_2$Te$_6$ except the ML 
because the out-of-plane magnetic anisotropy would open a spin-wave gap and thus suppress magnetic fluctuations so that
long-range FM order could be stabilized at finite temperature.  
In the visible frequency range, large Kerr rotations up to $\sim$1.0{\degree} 
in these materials are predicted and they are comparable
to that observed in famous MO materials such as PtMnSb and
Y$_3$Fe$_5$O$_{12}$. Moreover, they are $\sim$100 times larger than
that of 3$d$ transition metal MLs deposited on Au surfaces. 
Faraday rotation angles in these 2D materials are also large, being up 
to $\sim$120 deg/$\mu$m, and are thus comparable to the best-known MO semiconductor 
Bi$_3$Fe$_5$O$_{12}$. These findings thus suggest that with large MAE and MO effects, 
atomically thin Cr$_2$Ge$_2$Te$_6$ films would
have potential applications in novel magnetic, MO and spintronic nanodevices.
\end{abstract}

\maketitle


\section{Introduction}
Two-dimensional (2D) materials are layer substances with a thickness of one or few atomic or molecular
monolayers (MLs). The vibrant field of 2D materials was triggered by the first isolation of graphene,
a single atomic layer of graphite with carbon atoms arranged in a 2D honeycomb lattice, through 
mechanical exfoliation of graphite and also the discovery of its extraordinary transport properties 
in 2004.~\cite{novoselov2004} The recent intensive interest in graphene has also stimulated research
efforts to fabricate and investigate other 2D materials.~\cite{miro2014,bhimanapati2015} 
Many of these 2D materials exhibit
a variety of fascinating properties not seen in their bulk counterparts due to, e.g., symmetry breaking
and quantum confinement. For example, a semiconductor MoS$_2$ crystal was found to exhibit an indirect
to direct bandgap transition when thinned down to a ML.~\cite{mak2010} Importantly, the broken
inversion symmetry makes MLs of MoS$_2$ and other transition metal dichalcogenides exhibit novel
properties of fundamental and technological interest such as spin-valley coupling~\cite{xiao2012}, 
piezoelectricity~\cite{duerloo2012} and second-harmonic generation~\cite{wang2015}.

More recently, the family of 2D materials has also been extended to 2D magnets, crucial for
the development of magnetic nanodevices and spintronic applications. For example, realization 
of 2D magnets by doping transition metals into nonmagnetic parent monolayers has been 
predicted by {\it ab initio} density functional calculations and also 
attempted experimentally by numerous research groups~\cite{bhimanapati2015}. 
Hole-doping into a GaSe ML which possesses a unique
van Hove singularity near the top of its valance band, was recently predicted to induce
tunable ferromagnetism and magneto-optical effects~\cite{gao2015,feng2016}.   
Experimental and theoretical evidence for the ferromagnetic (FM) MoS$_2$ ML at
the MoS$_2$/CdS interface was also reported.~\cite{tan2016} Several layered magnetic compounds
have recently been investigated theoretically to determine whether their structure and magnetism
can be retained down to ML thickness.\cite{li2014,zhang2015b,sivadas2015,lin2016} 
These intensive research efforts finally led two groups to successfully thin bulk ferromagnets CrI$_3$ and
Cr$_2$Ge$_2$Te$_6$ down to few-layer ultrathin films and to observe the intrinsic ferromagnetism retained
in one and two monolayers, respectively.\cite{huang2017,gong2017} This successful fabrication
of 2D magnets would allow us to study 2D magnetism and exotic magnetic phenomena,
and also offers exciting prospects for advanced technological applications.

In this paper, we focus on two relativity-induced properties of
atomically thin FM Cr$_2$Ge$_2$Te$_6$ films\cite{gong2017}, namely, 
magnetic anisotropy energy (MAE) and magnet-optical (MO) effects.
For comparison, we also consider bulk Cr$_2$Ge$_2$Te$_6$, which is interesting 
on its own right. Bulk Cr$_2$Ge$_2$Te$_6$ belongs to the rare class of intrinsic FM 
semiconductors~\cite{carteaux1995} and has recently been attracting renewed interest 
because of its layered, quasi-2D FM structure with weak van der Waals bonds. 
For example, its magnetic anisotropy energy was recently measured.~\cite{zhang2016}
It was also used as a substrate for epitaxial growth of topological insulator Bi$_2$Te$_3$ 
to realize quantum anomalous Hall effect~\cite{ji2013}. 

Magnetic anisotropy energy refers to the total energy difference between
the easy- and hard-axis magnetizations, {\it i.e.}, the energy required to
rotate the magnetization from the easy to the hard direction.
Together with the magnetization and magnetic ordering temperature, MAE is one of the three 
important parameters that characterizes a FM material. Furthermore,
MAE is especially important for 2D magnetic materials such as few-layer Cr$_2$Ge$_2$Te$_6$ structures
because it helps to lift the Mermin-Wangner restriction~\cite{Mermin66}, thus allowing the
long-range ferromagnetic order to survive at finite temperature even in the 
ML limit~\cite{huang2017}. Moreover, 2D magnets with a significant MAE would 
find spintronic applications such as high density magnetic memory and data-storage devices. 
MAE consists of two contributions, namely, the magnetocrystalline anisotropy 
energy (C-MAE) due to the effect of relativistic spin-orbit coupling (SOC) 
on the electronic band structure, and also the magnetic dipolar (shape) anisotropy energy (D-MAE)
due to the magnetostatic interaction among the magnetic moments~\cite{Guo91b,Tung07}.
Note that the D-MAE always prefers an in-plane magnetization.~\cite{Guo91b}
Although the D-MAE in isotropic bulk materials is generally negligibly small, the D-MAE could become
significant in low-dimensional materials such as magnetic monolayers~\cite{Guo91b}
and atomic chains~\cite{Tung07}. Despite of the recent intensive interest in 
few-layer Cr$_2$Ge$_2$Te$_6$,~\cite{li2014,zhang2015b,sivadas2015,lin2016,gong2017,tian2016,xing2017} 
no investigation into their MAE has been reported, although an experimental 
measurement on MAE~\cite{zhang2016}  and a theoretical
calculation of C-MAE~\cite{gong2017} of bulk Cr$_2$Ge$_2$Te$_6$ were recently reported.  
Therefore, in this paper we carry out a systematic {\it ab initio} density functional investigation
into the MAE as well as other magnetic properties of bulk and few-layer Cr$_2$Ge$_2$Te$_6$. 
Furthermore, both C-MAE and D-MAE are calculated here.
Importantly, as will be reported in Sec. III, taking the D-MAE into account
results in an in-plane magnetization for the Cr$_2$Ge$_2$Te$_6$ ML
while all other considered Cr$_2$Ge$_2$Te$_6$ materials have the out-of-plane magnetic
anisotropy. Thus, this interesting finding could explain why the
ferromagnetism was not observed in the Cr$_2$Ge$_2$Te$_6$ ML.~\cite{gong2017}
 
Magneto-optical Kerr effect (MOKE) and Faraday effect (MOFE) are two well-known MO effects.~\cite{Oppeneer2001,antonov2004} 
When a linearly polarized light beam hits a magnetic material, the polarization vector 
of the reflected and transmitted 
light beams rotates. The former and latter are known as Kerr and Faraday effects, respectively.
Faraday effect has attracted much less attention than Kerr effect simply because light can only transmit
through ultrathin films. In contrast, MOKE has been widely used to study magnetic and electronic properties 
of solids including surfaces and films.~\cite{antonov2004} Furthermore, magnetic materials with large MOKE 
would find valuable MO storage and sensor applications~\cite{mansuripur95,castera96}, 
and hence have been continuously searched for in recent decades.
The recent development of 2D magnetic materials~\cite{huang2017,gong2017} offers exciting possibilities of scaling 
the MO storage and sensing devices to the few-nanometer scale. However, detection and measurement
of magnetism in these 2D materials will be difficult by using traditional methods such as superconducting
quantum interference device (SQUID) magnetometer because ultrahigh sensitivity would be required 
for these ultrathin films.
In this regard, MOKE and MOFE are a powerful, nonevasive probe of spontaneous magnetization 
in these 2D materials. Indeed, long-range ferromagnetic orders in both Cr$_2$Ge$_2$Te$_6$
and CrI$_3$ atomically thin films were detected by using the MOKE technique.~\cite{gong2017,huang2017,xing2017}
In this paper, therefore, we perform a systematic {\it ab initio} density functional study of
the magneto-optical and optical properties of bulk and atomically thin films of Cr$_2$Ge$_2$Te$_6$. 
Indeed, we find that both the bulk and the ultrathin films would exhibit large MOKE with Kerr rotation
angles being as large as $\sim$1.0 {\degree}. Furthermore, the calculated Faraday rotation angles
of the thin films are also large, being $\sim$120 deg/$\mu$m.
Therefore,  bulk and ultrathin films Cr$_2$Ge$_2$Te$_6$ are promising magneto-optical materials
for high-density MO recordings and magnetic nanosensors.

\section{STRUCTURES AND METHODS}

In this paper, we study the electronic, magnetic, and magneto-optical properties of few-layer and 
bulk Cr$_2$Ge$_2$Te$_6$ structures. Bulk Cr$_2$Ge$_2$Te$_6$ forms a layered structure with 
MLs separated by the so-called van der Waals (vdW) gap [Fig. 1(a)].\cite{carteaux1995}  
Each Cr$_2$Ge$_2$Te$_6$ ML consists of two AB-stacked compact hexagonal Te planes 
with one Ge dimer lying vertically at one of every three hexagon centers [Fig. 1(c)] and two Cr atoms
occupying two of every three Te octohedron centers [Figs 1(a), (c) and (d)]. These layers are ABC stacked
[Fig. 1(a)], thus resulting in a rhombohedral $R\bar{3}$ symmetry with one chemical formula unit (f.u.)
per unit cell. The experimental lattice constants are  $a=b=c=7.916$ \AA.~\cite{carteaux1995}.
This structure can also be regarded as an ABC-stacked hexagonal crystal cell with 
experimental lattice constants $a=b=6.828$ \AA$ $ and $c=20.5619$ \AA.~\cite{carteaux1995} 
As explained in the next paragraph, we adopt the experimental rhombohedral unit cell in the bulk calculations,
and for the few-layer Cr$_2$Ge$_2$Te$_6$ structures, the hexagonal unit cell with the experimental bulk 
lattice constants and atomic positions [Fig. 1(c)]. For bilayer (BL) and trilayer (TL)
Cr$_2$Ge$_2$Te$_6$ structures, we consider the AB and ABC stackings, respectively,
as observed in bulk Cr$_2$Ge$_2$Te$_6$ [Fig. 1(d)].
The few-layer structures are modelled by using the slab-superlattice approach with the separations
between MLs being at least 15 \AA.

\begin{figure}[htb]
\begin{center}
\includegraphics[width=7.5cm]{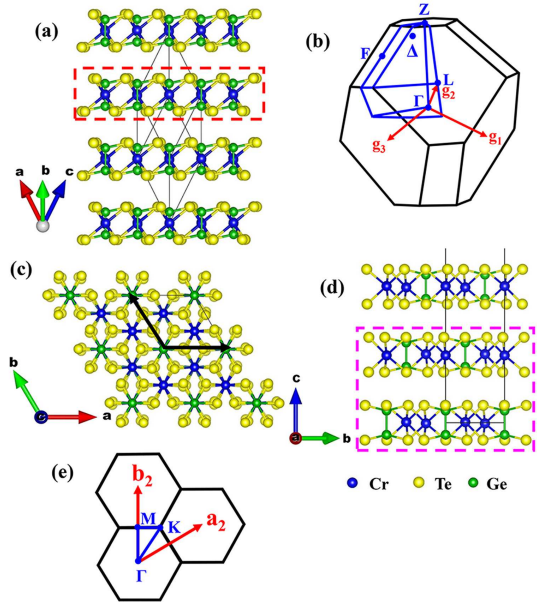}
\end{center}
\caption{(a) Crystalline structure of bulk rhombohedral Cr$_2$Ge$_2$Te$_6$ with the red dashed rectangle 
indicating one monolayer (ML) Cr$_2$Ge$_2$Te$_6$, and (b) the corresponding Brillouin zone with its irreducible
wedge indicated by blue lines.
(c) Top and (d) side views of trilayer (TL) Cr$_2$Ge$_2$Te$_6$ structure in the ABC stacking. 
In (d), the magenta dashed rectangle indicates a bilayer (BL) unit cell.
(e) Two-dimensional hexagonal Brillouin zone for ML, BL and TL  Cr$_2$Ge$_2$Te$_6$
with its irreducible wedge indicated by blue lines.} 
\end{figure}

{\it Ab initio} calculations are performed based on density functional theory.  
The exchange-correlation interaction is treated with the generalized gradient approximation 
(GGA) parameterized by Perdew-Burke-Ernzerhof formula~\cite{perdew1996generalized}. 
To improve the description of on-site Coulomb 
interaction between Cr 3{\it d} electrons, we adopt the GGA+U scheme.\cite{dudarev98} 
It was reported in Ref. \onlinecite{gong2017} that a physically appropriate 
value of the effective onsite Coulomb energy $U$
should be within the range of $0.2 < U < 1.7$ eV. Therefore, here we use $U=1.0$ eV (see also the 
supplementary note in the supplementary material (SM)\cite{SM} for explanations). 
The accurate projector-augmented wave~\cite{kresse1999ultrasoft} method, as implemented in the Vienna ab initio 
Simulation Package (VASP)~\cite{kresse1996efficient,kresse1996efficiency}, is used.  A large plane wave cutoff energy of 450 eV 
is used throughout. For the Brillouin zone integrations, $k$-point meshes of 16 $\times$ 16 $\times$ 16 and 20 $\times$ 20 $\times$ 1 
are used for bulk and few-layer Cr$_2$Ge$_2$Te$_6$, respectively.
The structural optimization within the GGA+U scheme for bulk Cr$_2$Ge$_2$Te$_6$ results in 
lattice constants $a = 6.931$ \AA$ $ and $c = 22.695 $ \AA$ $. This theoretical $c$ 
is more than 10 \% larger than the experimental $c$, although the calculated $a$ is only 1.5 \% too large. 
This is because the GGA+U tends to largely overestimate the vdW gaps in layered materials.   
To account for the vdW dispersion interactions, we further perform the structural
optimization with the GGA+U plus vdW-density functional of Langreth and co-workers~\cite{Lee10} as
implemented in the VASP, and obtain $a = 6.905$ \AA$ $ and $c = 20.074 $ \AA$ $.
The discrepancy in $c$ between the calculation and experiment~\cite{carteaux1995}
is much reduced to -2.3 \% but the calculated $a$ is still more than 1.0 \% too large.
The GGA+U structural optimization for ML  Cr$_2$Ge$_2$Te$_6$ results in $a = 6.925$ \AA$ $,
which deviates only slightly from the theoretical bulk $a$ (by less than 0.1 \%).
This shows that structural relaxations in ML Cr$_2$Ge$_2$Te$_6$ is much smaller than
the errors in the GGA+U and GGA+U+vdW correction schemes. 
Therefore, we use the experimentally determined bulk atomic structure in all the subsequent calculations.

\begin{figure}[htb]
\begin{center}
\includegraphics[width=7.5cm]{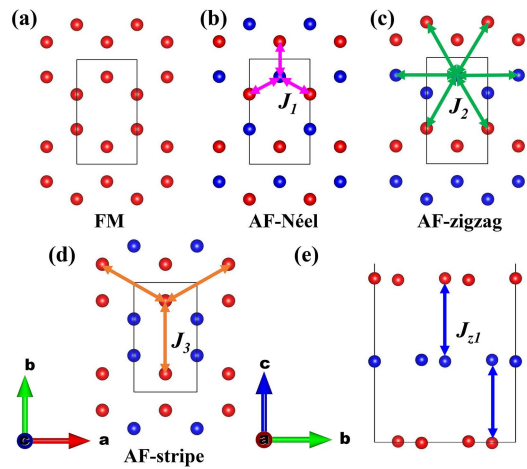}
\end{center}
\caption{Four considered intralayer magnetic configurations: (a) FM, 
(b) AF-N$\rm\acute{e}$el, (c) AF-zigzag, (d) AF-stripe types. Only magnetic Cr atoms are shown
with red (blue) balls indicating up (down) spins. Intralayer exchange coupling parameters
$J_1$, $J_2$, and $J_3$ are indicated by magenta, green, and orange arrows, respectively.
(e) Interlayer AF configuration with the interlayer exchange coupling parameter $J_{z1}$ 
indicated by blue lines.}	
\end{figure}

To understand the magnetism and also estimate the magnetic ordering temperature ({\it {T$_c$}}) 
for bulk and few-layer Cr$_2$Ge$_2$Te$_6$, we determine the exchange coupling parameters 
by mapping the calculated total energies of different 
magnetic configurations onto the classical Heisenberg Hamiltonian,
$ 
{\it E=E_0-\sum_{i,j}J_{ij}\hat{\mathbf{e}}_i\ {\bm\cdot}\ \hat{\mathbf{e}}_j},
$ 
where $E_0$ donates the nonmagnetic ground state energy; $J_{ij}$ the exchange coupling parameter between sites $i$ and $j$;
$\hat{\mathbf{e}}_i$, the unit vector representing the direction of the magnetic moment
on site $\it i$. Specifically, the total energies of the four intralayer magnetic configurations
for one Cr$_2$Ge$_2$Te$_6$ layer can be expressed as a set of 4 linear equations of $J_1$, $J_2$ and $J_3$: 
$E_{FM}=E_0-3J_1-6J_2-3J_3$, $E_{AF-Neel}=E_0+3J_1-6J_2+3J_3$, $E_{AF-zigzag}=E_0-J_1+2J_2+3J_3$
and $E_{AF-stripe}=E_0+J_1+2J_2-3J_3$. Given the calculated total energies, one can solve
this set of linear equations to obtain the $J_1$, $J_2$ and $J_3$ values.
Similarly, given the calculated $E_{FM}$ and $E_{AF-interlayer}$, one can obtain
the $J_{z1}$ value.
 
As described in Sec. I, magnetic anisotropy energy (MAE) consists of two parts, namely,
magnetocrystalline anisotropy energy (C-MAE) and magnetic dipolar anisotropy energy (D-MAE).
To determine C-MAE, we first perform two self-consistent relativistic electronic structure
calculations for the in-plane and out-of-plane magnetizations, and then 
obtain the C-MAE as the total energy difference between the two calculations. 
Highly dense $k$-point meshes of 25 $\times$ 25 $\times$ 25 and 30 $\times$ 30 $\times$ 1 
are used for bulk and few-layer Cr$_2$Ge$_2$Te$_6$, respectively. Test calculations
using different $k$-point meshes show that thus-obtained C-MAE converges within 1 \%. 
For a FM system, the magnetic dipolar energy $E_d$ in atomic Rydberg units is given by~\cite{Guo91b,Tung07}
\begin{equation}
E_d=\sum_{qq'}\frac{2m_qm_{q'}}{c^2}M_{qq'},
\end{equation}
where the so-called  magnetic dipolar Madelung constant
\begin{equation}
M_{qq'} = {\sum_{\bf R}}^{'}\frac{1}{\mid \bf R+q+q^{'}\mid^{3}}\{1-3\frac{[(\bf R+q+q^{'})\cdot \hat{m_q}]^2}{\mid \bf R+q+q^{'}\mid^{2}}\}
\end{equation} 
which is evaluated by Ewald's lattice summation technique.~\cite{Ewald21} 
{\bf R} are the lattice vectors and {\bf q} are the atomic position vectors in the unit cell. 
The speed of light $c=274.072$ and $m_q$ is the atomic magnetic moment
(in units of $\mu_B$) on site $q$. Thus, given the calculated magnetic moments,
the D-MAE is obtained as the difference in $E_d$ between the in-plane and out-of-plane magnetizations.
In a 2D material, all the {\bf R} and {\bf q} are in-plane. Thus, the second term in Eq. (2) 
would be zero for the out-of-plane magnetization, thereby resulting in the positive $M_{qq'}$ while
for an in-plane magnetization the $M_{qq'}$ are negative. For example, 
in ML Cr$_2$Ge$_2$Te$_6$, the calculated $M_{11}$ ($M_{12}$) is 11.0246 (23.0628) $a^{-3}$ 
for the out-of-plane magnetization and is -5.5123 (-11.5314) $a^{-3}$ for an in-plane magnetization.
Therefore, the D-MAE always prefers an in-plane magnetization in a 2D material.
This is purely a geometric effect and consequently the D-MAE is also known as the
magnetic shape anisotropy energy.  

For a FM solid possessing at least a trigonal symmetry with the magnetization along the
rotational $z$-axis, the optical conductivity tensor can be reduced in the form, 
\begin{equation}
\sigma=
\begin{pmatrix}
 \sigma_{xx} & \sigma_{xy}   & 0 \\
 -\sigma_{xy}& \sigma_{xx} & 0 \\
 0 & 0  & \sigma_{zz}
\end{pmatrix}
\end{equation}
We calculate the three independent elements of the conductivity tensor using the Kubo formula 
within the linear response theory.~\cite{wang1974band,Oppeneer92,feng2015large} 
Here the adsorptive parts of these elements, i.e., the real diagonal and imaginary off-diagonal elements, 
are given by,
\begin{equation}
\sigma_{aa}^{1} (\omega) = \frac{\pi e^2}{\hbar\omega m^2}
\sum_{i,j}\int_{BZ}\frac{d{\bf k}}{(2\pi)^3}|p_{ij}^{a}|^{2}
\delta(\epsilon_{{\bf k}j}-\epsilon_{{\bf k}i}-\hbar\omega),
\end{equation}
\begin{equation}
\sigma_{xy}^{2} (\omega) = \frac{\pi e^2}{\hbar\omega m^2}
\sum_{i,j}\int_{BZ}\frac{d{\bf k}}{(2\pi)^3}\text{Im}[p_{ij}^{x}p_{ji}^{y}]
\delta(\epsilon_{{\bf k}j}-\epsilon_{{\bf k}i}-\hbar\omega),
\end{equation}
where $\hbar$$\omega$ is the photon energy, and $\epsilon_{{\bf k}i}$ is the $i$th band energy 
at point ${\bf k}$. Summations $i$ and $j$ are over the occupied and unoccupied bands, respectively. 
Dipole matrix elements $p_{ij}^{a} = \langle\textbf{k}\emph{j}|\hat{p}_{a}|\textbf{k}i\rangle$
where $\hat{p}_a$ denotes Cartesian component $a$ of the dipole operator,
are obtained from the band structures within the PAW formalism\cite{Adolph01},
as implemented in the VASP package. The integration over the Brillouin zone is carried out
by using the linear tetrahedron method (see ~\cite{Temmerman89} and references therein).
The dispersive parts of the optical conductivity elements can be obtained from the corresponding 
absorptive parts using the Kramers-Kroing relations
\begin{equation}
\sigma_{aa}^{2} (\omega) = -\frac{2\omega}{\pi }P \int _{0}^{\infty }\frac{\sigma_{aa}^{1}(\omega ')}{\omega^{'2}-\omega ^{2}}d\omega ^{'},
\end{equation}
\begin{equation}
\sigma_{xy}^{1} (\omega) = \frac{2}{\pi }P \int _{0}^{\infty }\frac{\omega^{'}\sigma_{xy}^{2}(\omega ')}{\omega^{'2}-\omega ^{2}}d\omega ^{'},
\end{equation}
where $P$ donates the principle value. 
To take the finite quasiparticle lifetime effects into account, we convolute all the optical conductivity
spectra with a Lorentzian of line width $\Gamma$. For layered vdW materials such as graphite,
$\Gamma$ is about 0.2 eV (see, e.g., Figs. 1(a) and 1(b) in Ref. [\onlinecite{Guo04}]), which is thus
used in this paper. 

We should note that the optical conductivity calculated using Eqs. (4) and (5) is based on the independent-particle approximation, 
i.e., the many-body effects, namely, the quasi-particle self-energy corrections and excitonic effects, 
are neglected. These many-body effects on the optical properties of 2D systems such as MoS$_2$ and SiC monolayers \cite{Qiu13,Hsueh11} 
are especially pronounced due to the reduced dimensionality. Nevertheless, here the self-energy corrections are taken into account 
by the scissors correction using the bandgaps from the hybrid functional HSE06 \cite{heyd2003j,heyd2006} calculations, 
as reported in the next section (Sec. III.B), and the quasi-particle lifetime effects are accounted for by
convoluting all the optical conductivity spectra with a Lorentzian, as mentioned above. 
Weak or moderate electron-hole interaction would merely enhance the peaks near the absorption edge \cite{Benedict98}. 
However, strong electron-hole interaction in, e.g., MoS$_2$ monolayer, could give rise to additional prominent excitonic 
peaks below the absorption edge \cite{Qiu13}. Nonetheless, in contrast to direct bandgap MoS$_2$ monolayer, 
all the Cr$_2$Ge$_2$Te$_6$ systems considered here have an indirect bandgap and hence one could expect no strong
excitonic effect on their optical and MO properties. 
Indeed, no excitonic transition peak was observed in ultrathin indirect band gap GaSe films in a very recent experiment \cite{Budweg17}. 

Here we consider the polar MOKE and MOFE. For a bulk magnetic material, the complex polar Kerr rotation angle is given by~\cite{guo1994,guo1995},
\begin{equation}
\theta _{K}+i\epsilon _{K}=\frac{-\sigma _{xy}}{\sigma _{xx}\sqrt{1+i(4\pi/\omega)\sigma _{xx}}}.
\end{equation}
For a magnetic thin film on a nonmagnetic substrate, however, the complex polar Kerr rotation angle is given by~\cite{feng2016,suzuki1992}
\begin{equation}
\theta _{K}+i\epsilon _{K}=i\frac{2\omega d}{c}\frac{\sigma _{xy}}{\sigma_{xx}^{s}} = \frac{8\pi d}{c}\frac{\sigma _{xy}}{(1-\varepsilon_{xx}^{s})}
\end{equation}
where $c$ stands for the speed of light in vacuum; $d$ the thickness of the magnetic layer; $\varepsilon_{xx}^{s}$ ($\sigma_{xx}^s$) 
the diagonal part of the dielectric constant (optical conductivity) of the substrate. Experimentally, 
atomically thin Cr$_2$Ge$_2$Te$_6$ films were prepared on SiO$_2$/Si.~\cite{gong2017}
Thus, the optical dielectric constant of bulk SiO$_2$ ($\sim$3.9) is used as $\varepsilon_{xx}^s$.
Similarly, the complex Faraday rotation angle for a thin film can be written as~\cite{ravindran1999}
\begin{equation}
\theta _{F}+i\epsilon _{F}=\frac{\omega d}{2c}(n_{+}-n_{-}),
\end{equation}
where $n_+$ and $n_-$ represent the refractive indices for left- and right-handed polarized lights, respectively, 
and are related to the corresponding dielectric function (or optical conductivity via expressions 
$n_{\pm }^{2}=\varepsilon_{\pm}=1+{\frac{4\pi i}{\omega}}\sigma _{\pm}= 1+{\frac{4\pi i}{\omega}}(\sigma _{xx}\pm i \sigma _{xy})$.
Here the real parts of the optical conductivity $\sigma _{\pm}$ can be written as
\begin{equation}
\sigma_{\pm}^{1} (\omega) = \frac{\pi e^2}{\hbar\omega m^2}
\sum_{i,j}\int_{BZ}\frac{d{\bf k}}{(2\pi)^3}|\Pi_{ij}^{\pm}|^{2}
\delta(\epsilon_{{\bf k}j}-\epsilon_{{\bf k}i}-\hbar\omega),
\end{equation}
where $\Pi_{ij}^{\pm} = \langle\textbf{k}\emph{j}|\frac{1}{\sqrt{2}}(\hat{p}_{x}\pm i\hat{p}_{y}|\textbf{k}i\rangle$. 
Clearly, $\sigma _{xy} = \frac{1}{2i}(\sigma _{+}-\sigma _{-})$, and this shows that $\sigma _{xy}$ 
would be nonzero only if $\sigma _{+}$ and $\sigma _{-}$ are different. 
In other words, magnetic circular dichroism is the fundamental cause of the
nonzero $\sigma _{xy}$ and hence the magneto-optical effects.

\section{RESULTS AND DISCUSSION }
\subsection{Magnetic moments and exchange coupling}
We consider four intralayer magnetic configurations for each Cr$_2$Ge$_2$Te$_6$ ML comprising the FM as well as three antiferromagnetic 
(AF) structures, as labeled AF-N$\rm \acute{e}$el, AF-zigzag, and AF-stripe in Fig. 2. The FM state is found to be
the ground state in all the considered materials. The lowest energy AF state is the AF-zigzag which
is however more than 0.05 eV/f.u. higher in energy than the FM state in all the systems considered. 
For the bilayer, trilayer and bulk systems, we also consider the interlayer AF state [see Fig. 2(e)].
We find that the interlayer AF state is only slightly (i.e., a couple of meV/f.u.) 
above the interlayer FM state. Calculated spin and orbital magnetic moments of Cr$_2$Ge$_2$Te$_6$
in the FM state are listed in Table I. The Cr spin magnetic moment in all the structures 
is $\sim$3.2 $\mu_B$, being in good agreement with the experimental value of $\sim$3.0 $\mu_B$\cite{ji2013}. 
This is consistent with three unpaired 
electrons in the Cr$^{3+} (d^3; t_{2g}^{3\uparrow})$ ionic configuration in these materials.
The calculated Cr orbital magnetic moment is negligibly small (Table I), further suggesting that
the Cr ions are in the $(d^{3}; t_{2g}^{3\uparrow})$ configuration.
Interestingly, the induced spin magnetic moment on the Te site is significant ($\sim$0.15 $\mu_B$) and also antiparallel
to that of the Cr atoms, although the spin moment on the Ge atom is five times smaller (Table I).  
This results in exactly 6.0  $\mu_B$/f.u., and is in good agreement with the measured bulk magnetization\cite{ji2013}.
The mechanism of ferromagnetism in all the Cr$_2$Ge$_2$Te$_6$ structures could be attributed to
the dominant FM superexchange coupling between half-filled Cr $t_{2g}$ and empty $e_g$ states 
via Te $p$ orbitals, against the AF direct exchange interactions of Cr $t_{2g}$ states.\cite{li2014} 
This is further supported by our finding of significant spin moments of Te atoms which are
antiparallel to the Cr spin moments (Table I). 

\begin{table*}[htbp]
\begin{center}
\caption{Total spin magnetic moment ($m_s^t$), atomic (averaged) spin ($m_s^{Cr}$, $m_s^{Ge}$, $m_s^{Te}$) and orbital 
($m_o^{Cr}$, $m_o^{Ge}$, $m_o^{Te}$) magnetic moments as well as band gap ($E_g$) of bulk and 
few-layer FM Cr$_2$Ge$_2$Te$_6$ (magnetization being perpendicular to the layers) calculated 
using the GGA+U scheme with the spin-orbit coupling included. 
Also listed are total magnetic anisotropy energy ($\Delta E_{ma}$), magnetocrystalline anisotropy energy 
($\Delta E_{b}$) and magnetic dipolar anisotropy energy ($\Delta E_{d})$. 
Positive $\Delta E_{ma}$ means that the out-of-plane magnetization is the easy axis.
For comparison, the available experimental $E_g$ and $\Delta E_{ma}$ values are also listed.
The bandgaps ($E_g^{HSE}$) calculated using the hybrid HSE06 functional are also included here.}
\begin{ruledtabular}
\begin{tabular}{ccccccccc}
structure  & $m_s^t$ & $m_s^{Cr}$ ($m_o^{Cr}$) & $m_s^{Ge}$ ($m_o^{Ge}$) & $m_s^{Te}$ ($m_o^{Te}$) & 
$\Delta E_{b}$ ($\Delta E_{d}$) & $\Delta E_{ma}$ & $E_g$ ($E_g^{exp}$)  & $E_g^{HSE}$  \\
        &  ($\mu_B$/f.u.) & ($\mu_B$/atom) & ($\mu_B$/atom) & ($\mu_B$/atom) & (meV/f.u.) & (meV/f.u.) & (eV) & (eV) \\
 \hline
  monolayer & 6.00 & 3.23 (0.003) & 0.024 (0.001) & -0.144 (-0.003)& 0.107 (-0.153) &-0.05 & 0.23  & 0.56 \\
    bilayer & 6.00 & 3.23 (0.003) & 0.024 (0.001) & -0.145 (-0.003)& 0.274 (-0.153) & 0.12 & 0.13  & 0.44 \\
   trilayer & 6.00 & 3.23 (0.003) & 0.025 (0.001) & -0.145 (-0.001)& 0.297 (-0.153) & 0.14 & 0.09  & 0.41 \\
       bulk & 6.00 & 3.23 (0.003) & 0.025 (0.001) & -0.146 (-0.003)& 0.471 (-0.067) & 0.40 (0.05\footnotemark[1]) & 0.04 (0.74\footnotemark[2]) & 0.33 
\end{tabular}
\end{ruledtabular}
\footnotemark[1]{Ref.~\onlinecite{zhang2016}.}
\footnotemark[2]{Ref.~\onlinecite{ji2013}.}
\end{center}
\end{table*}

Using the calculated total energies of the magnetic configurations considered here, 
we estimate the intralayer first-, second- and third-neighbor exchange coupling parameters 
($J_1$, $J_2$, $J_3$) as well as the first-neighbor interlayer coupling constant ($J_{z1}$) 
by mapping the total energies to the effective Heisenberg Hamiltonian model as described 
above in Section II.  The estimated exchange coupling parameters are listed in Table II. 
The large, positive $J_1$ values in all the structures show that the nearest-neighbor Cr-Cr coupling 
is strongly ferromagnetic. In contrast, the second nearest neighbor coupling ($J_2$) 
is about 20 times weaker and antiferromagnetic. Interestingly, calculated $J_3$ values 
are three times larger than $J_2$ although they are about ten times smaller than $J_1$. 
The calculated interlayer exchange coupling parameter ($J_{z1}$) is about as large as $J_2$
but ferromagnetic. Table I shows that all the exchange interaction parameters increase
monotonically with the increasing number of layers. This suggests that
the magnetism is strengthened as one moves from the ML to the BL and eventually to the bulk.
Our $J_1$ value agree quite well with that of the previous {\it ab initio} calculation ($\sim 12$ meV)
reported in Ref. \cite{li2014}. Nonetheless, all the present $J$ values appear to be
about twice as large as that of the {\it ab initio} calculation of Ref. \cite{gong2017},
although the trends are very similar. This discrepancy in the calculated J values 
between Ref. [\onlinecite{gong2017}] and the present calculation could be attributed to 
the fact that the LSDA+U is used in Ref. [\onlinecite{gong2017}] while the GGA+U is exploited in this paper. 
This is because the LSDA tends to underestimate the tendency of magnetism.
Note that for comparison of the previous {\it ab initio} 
calculations\cite{li2014,gong2017} with the present results, the $J$ values reported in 
Refs. \cite{li2014} and \cite{gong2017} should be multiplied by a factor of $S^2 = 9/4$.

\subsection{Magnetic anisotropy energy and ferromagnetic transition temperature}

\begin{table}[htbp]
\begin{center}
\caption{
Intralayer first-, second- and third-neighbor exchange coupling parameters
($J_1$, $J_2$, $J_3$) as well as interlayer first-neighbor exchange coupling constant ($J_{z1}$) 
in monolayer (ML), bilayer (BL), trilayer (TL) and bulk Cr$_2$Ge$_2$Te$_6$,
derived from the calculated total energies for various magnetic configurations (see the text). 
Ferromagnetic transition temperatures estimated using the derived $J_1$ values within the 
mean-field approximation ($T_c^m$) and within the mean-field approximation plus spin wave gap correction 
($T_c$) (see the text) are also included. For comparison, the available experimental 
$T_c$ values~\cite{gong2017} are also listed in brackets.}
\begin{ruledtabular}
\begin{tabular}{ccccccc}
system&  $J_1$ &$J_2$ & $J_3$ &$J_{z1}$ & $T_c^m$ & $T_c$ ($T_c^{exp}$) \\
         & (meV) &(meV) & (meV) &  (meV) & (K) & (K) \\ \hline
 ML &12.74 & -0.09 & 0.29 & -    & 149 & 0    \\
 BL &14.26 & -0.47 & 1.07 & 0.40 & 169 & 71 (28\footnotemark[1]) \\
 TL &14.75 & -0.57 & 1.29 & 0.74 & 176 & 76 (35\footnotemark[1])  \\
bulk &15.74 & -0.78 & 1.74 & 1.08 & 189 & 99 (66\footnotemark[1]) \\
\end{tabular}
\end{ruledtabular}
\footnotemark[1]{Ref.~\onlinecite{gong2017}.}
\end{center}
\end{table}

As described in Sec. I, magnetic anisotropy energy consists of
magnetocrystalline anisotropy energy ($\Delta E_{b}$) due to the SOC effect
on the band structure, and magnetic dipolar anisotropy energy 
($\Delta E_{d}$) due to the magnetostatic interaction among the magnetic dipoles~\cite{Guo91b,Tung07}.
Although the D-MAE in bulk materials is generally negligibly small, the D-MAE could become
significant in low-dimensional materials such as magnetic monolayers~\cite{Guo91b} 
and atomic chains~\cite{Tung07}. Therefore, we calculate both C-MAE and D-MAE and 
Table I lists all calculated C-MAE ($\Delta E_{b}$), D-MAE ($\Delta E_{d}$) and MAE ($\Delta E_{ma}$).
Indeed $\Delta E_{d}$ is comparable to $\Delta E_{b}$ in
the ML and BL (Table I). Note that the D-MAE always prefers an in-plane magnetization\cite{Guo91b}, and it
thus competes with the C-MAE which prefers the out-of-plane magnetic anisotropy 
in all the systems considered here. In fact, the magnitude of the D-MAE in the ML is 
larger than that of the C-MAE and this results in an in-plane magnetization. 

Table I shows clearly that calculated MAEs for all the investigated materials except the ML,
are large, and more importantly, prefer the out-of-plane magnetization (i.e., having a positive MAE value).
In particular, they are two orders of magnitude larger than that ($\sim 5$ $\mu$eV/atom) of
elemental ferromagnets Fe and Ni~\cite{Guo91a}. They are also comparable to that of layered heavy element 
magnetic alloys such as FePt and CoPt~\cite{Oppeneer98} which have the largest MAEs 
among magnetic transition metal alloys. 
Because of the appreciable out-of-plane anisotropy, all the Cr$_2$Ge$_2$Te$_6$ materials except the ML,
could have possible applications in high density magnetic data storage.

Since $J_1$ is dominant (Table II), we estimate FM ordering temperatures within
the mean-field approximation, i.e., $k_BT_c^m = \frac{1}{3}zJ_1$ where $z$ is the number of
first-near neighbor Cr atoms\cite{Halilov98} and $z = 3$ in the present cases (Fig. 2).
Table II shows that $T_c^m$ increases from 149 K to 189 K as one goes from the ML 
to the bulk, a trend being consistent with the monotonical increase in the $J$ values mentioned above.
This trend also agrees with the experimental result\cite{gong2017}, although the $T_c^m$ values
are several times too large (see Table II). The much larger obtained $T_c^m$ values
could be caused by the mean-field approximation and perhaps also by our neglect of oscillatory (see Table II)
longer distance exchange couplings.\cite{Halilov97}. 
The mean-field approximation works well for bulk magnets with a high number
of neighboring magnetic atoms such as Fe and Ni metals~\cite{Halilov98}. Nevertheless, it would substantially
overestimate the $T_c$ for 2D materials with a much reduced coordination number 
such as Fe and Co MLs because it neglects transverse spin fluctuations~\cite{Pajda00,Irkhin99}. 
More importantly, the mean-field approximation violates Mermin-Wagner theorem~\cite{Mermin66} which says 
that long-range magnetic order at finite temperature cannot exist in an isotropic 2D Heisenberg magnet. 
However, the out-of-plane anisotropy can stabilize long-range magnetic orders at finite temperature by
opening a gap in the spin wave spectrum which suppresses transverse spin fluctuations.~\cite{Pajda00,Irkhin99} 
Note that an in-plane magnetic anisotropy would not produce a gap in the spin wave spectrum. 
Thus, a better approach is the spin wave theory with random-phase approximation  
which takes the out-of-plane anisotropy into account and hence meets Mermin-Wagner theorem~\cite{Pajda00}. 
The $T_c$ values estimated within the spin wave theory with random-phase approximation
for Fe and Co MLs are about a factor of 3 smaller than that from the mean-field approximation~\cite{Pajda00}.
It was reported that taking the out-of-plane anisotropy into account within self-consistent spin wave theory would lead to
a renormalization of $T_c^m$ by a factor of $log(k_BTc^m/\Delta E_{ma})$.~\cite{Irkhin99}
Based on the calculated MAE values, we come up with a set of renormalized $T_c$ values 
as listed in Table II. The renormalized $T_c$ values agree reasonably well with the 
corresponding experimental values (Table II). The $T_c$ for the ML is zero
because it has an in-plane easy axis of magnetization. This also explains why the long-range FM
ordering was not observed in the ML\cite{gong2017}.

\begin{figure}[htb]
\begin{center}
\includegraphics[width=8cm]{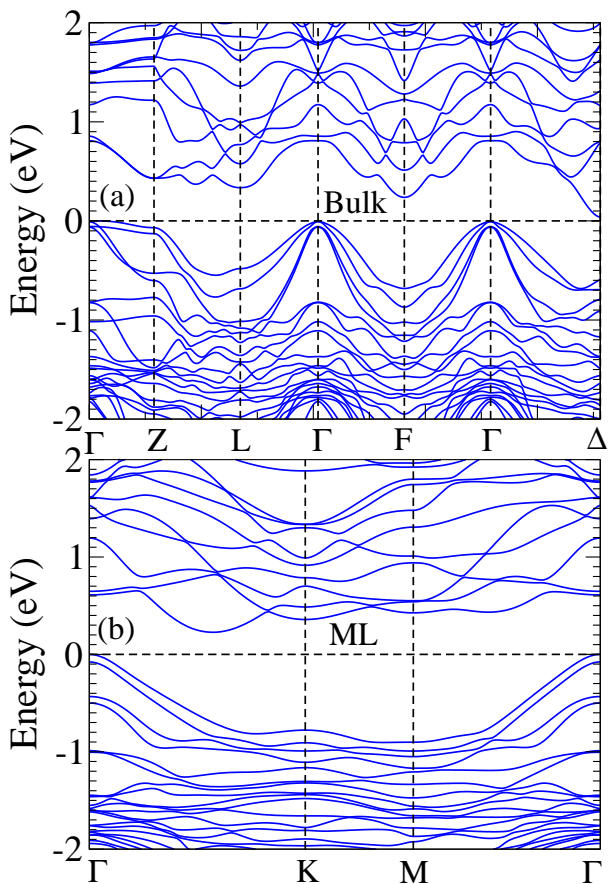}
\end{center}
\caption{Relativistic band structures of (a) bulk and (b) ML Cr$_2$Ge$_2$Te$_6$
in the FM state with the out-of-plane magnetization. 
Horizontal dashed lines denote the top of valance band.}
\end{figure}

\subsection{\label{sec:level2}Electronic structure}
In order to understand the electronic, magnetic and optical properties of the Cr$_2$Ge$_2$Te$_6$ materials,
we calculate their electronic band structures. Relativistic band structures of bulk and 
monolayer Cr$_2$Ge$_2$Te$_6$ are displayed in Fig. 3, while that of bilayer and trilayer Cr$_2$Ge$_2$Te$_6$ 
are shown in Fig. S1 in the SM~\cite{SM}. It is clear from Fig. 3 and Fig. S1 
that all the four structures 
are indirect bandgap semiconductors with the valence band maximum (VBM) being at the $\Gamma$ point. 
The conduction band minimum (CBM) in the Cr$_2$Ge$_2$Te$_6$ multilayers is located at somewhere 
along the $\Gamma$-K symmetry line (see Fig. 3 and Fig. S1)
while that of bulk Cr$_2$Ge$_2$Te$_6$ is located at a general $k$-point of (0.4444, -0.3704, 0.2222)$2\pi/a$.
Interestingly, calculated spin-polarized scalar-relativistic band structures (see Fig. S2 in ~\cite{SM}), 
indicate that both the CBM and VBM are of purely spin-up character.
The calculated bandgaps ($E_g$) are listed in Table I. It is seen that the theoretical bandgap
of bulk Cr$_2$Ge$_2$Te$_6$ is much smaller than the experimental value. This significant discrepancy
between experiment and theory is due to the well-known underestimation of the bandgaps by the GGA. 
In order to obtain accurate optical properties which are calculated from the band structure, 
we also perform the band structure calculations using the hybrid  
Heyd-Scuseria-Ernzerhof (HSE) functional.~\cite{heyd2003j,heyd2006}
The HSE band structures are displayed in Fig. S3 in the SM\cite{SM} and the HSE band gaps are listed in Table I. 
The HSE bandgap of bulk Cr$_2$Ge$_2$Te$_6$ is now comparable to the experimental value (Table I).
Therefore, the optical properties of all the Cr$_2$Ge$_2$Te$_6$ materials are calculated from 
the GGA band structures within the scissors correction scheme\cite{Levine91} using the differences
between the HSE and GGA bandgaps (Table I).

\begin{figure}[htb]
\begin{center}
\includegraphics[width=3.0in]{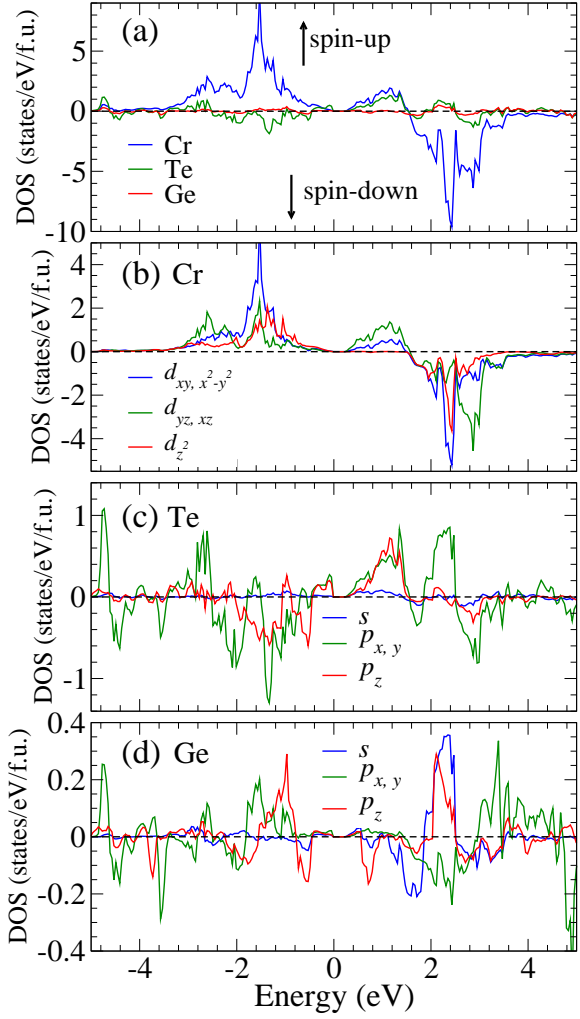}
\end{center}
\caption{Scalar-relativistic site-, orbital-, and spin-projected densities of states (DOS) of 
ferromagnetic bulk Cr$_2$Ge$_2$Te$_6$.  The top of the valence band is at 0 eV.}
\end{figure}

\begin{figure}[htb]
\begin{center}
\includegraphics[width=3.0in]{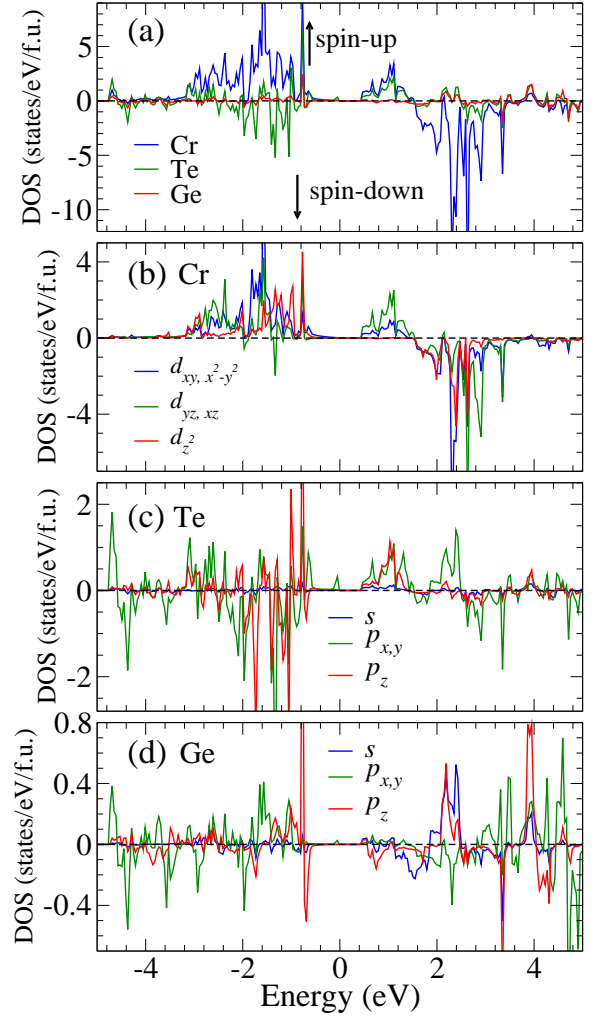}
\end{center}
\caption{Scalar-relativistic site-, orbital-, and spin-projected densities of states (DOS) of 
FM ML Cr$_2$Ge$_2$Te$_6$.}
\end{figure}

We also calculate total as well as site-, orbital-, and spin-projected densities of states (DOS) 
for all the Cr$_2$Ge$_2$Te$_6$ materials. In Fig. 4, we display site-, orbital- and spin-projected DOSs
of bulk Cr$_2$Ge$_2$Te$_6$. It can be seen that the upper valence band ranging from -4.0 eV to -0.3 eV
and also the lower conduction band ranging from 0.4 eV to 4.0 eV originate mainly from Cr $d$ orbitals 
with minor contributions from Te $p$ orbitals due to the hybridization between Cr $d$ and Te $p$ orbitals. 
The valence bands below these spin-up Cr $d$-dominant bands, are primarily derived from Te $p$ orbitals.
Furthermore, the upper valence band ranging from -3.3 eV to -0.3 eV consists of a broad peak of 
mainly spin-up Cr $d_{xy,x^2-y^2}$ orbitals. 
The lower conduction bands ranging from 0.4 eV to 1.5 eV is mainly made up of spin-up 
Cr $d_{xz,yz}$ orbitals. This suggests that the bandgap is created by
the crystal-field spliting of spin-up Cr $d_{xy,x^2-y^2}$ and $d_{xz,yz}$ bands.
Above this up to 3.6 eV, the conduction band consists of
a pronounced peak of spin-down Cr $d_{xy,x^2-y^2}$ and $d_{z^2}$ orbitals.  

Site-, orbital- and spin-projected DOSs of ML Cr$_2$Ge$_2$Te$_6$ are shown in Fig. 5.
Clearly, the features in the DOSs spectra of the ML are rather similar to that in the DOSs spectra
of bulk Cr$_2$Ge$_2$Te$_6$ (Fig. 4). The main difference appears to be that the contributions from the $p$ orbitals
of Te and Ge atoms are enhanced in the higher energy region. Another difference is the enhanced bandgap 
in the ML mainly due to the lack of interlayer interaction. 
Site-, orbital- and spin-projected DOSs of BL and TL Cr$_2$Ge$_2$Te$_6$ fall between 
that of ML and bulk Cr$_2$Ge$_2$Te$_6$ and thus are not shown here.

According to perturbation theory analysis, only the occupied and unoccupied Cr $d$ states near the Fermi level
which are coupled by the SOC, would significantly contribute to the magneto-crystalline anisotropy.~\cite{wang1993}
Moreover, the SOC matrix elements $\langle d_{xz}|H_{SO}|d_{yz} \rangle$ and $\langle d_{x^2-y^2}|H_{SO}|d_{xy} \rangle$
are found to prefer the out-of-plane anisotropy, while $\langle d_{x^2-y^2}|H_{SO}|d_{yz} \rangle$, $\langle d_{xy}|H_{SO}|d_{xz} \rangle$
and $\langle d_{z^2}|H_{SO}|d_{yz} \rangle$ favor an in-plane anisotropy.~\cite{Takayama}
The ratio of these matrix elements are $\langle d_{xz}|H_{SO}|d_{yz} \rangle^2$:$\langle d_{x^2-y^2}|H_{SO}|d_{xy} \rangle^2$:
$\langle d_{x^2-y^2}|H_{SO}|d_{yz}\rangle^2$:$\langle d_{xy}|H_{SO}|d_{xz}\rangle^2$
:$\langle d_{z^2}|H_{SO}|d_{yz}\rangle^2$ = $1:4:1:1:3$. 
Figures 4(b) and 5(b) show that in both bulk and ML Cr$_2$Ge$_2$Te$_6$, Cr $d_{z^2}$ DOS
is almost zero in the CBM region of 0.4$\sim$1.5 eV, and consequently, matrix 
element $\langle d_{z^2}|H_{SO}|d_{yz}\rangle$
would be negligibly small and thus hardly contribute to the C-MAE. In contrast, Cr $d_{xy,x^2-y^2}$ DOS
has prominent peaks in both VBM and CBM regions, and hence matrix element $\langle d_{x^2-y^2}|H_{SO}|d_{xy} \rangle$
would be large. This would give rise to a dominating contribution to the C-MAE. All these together would then lead to
a C-MAE which prefers the out-of-plane anisotropy in the Cr$_2$Ge$_2$Te$_6$ materials.     

\begin{figure}[htb]
\begin{center}
\includegraphics[width=8cm]{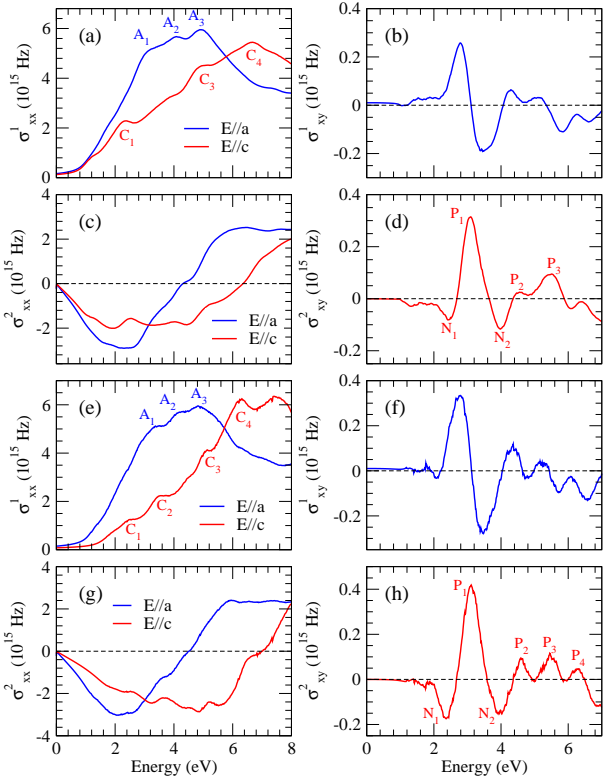}
\end{center}
\caption{Real (a) [(e)] diagonal and (b) [(f)] off-diagonal, imaginary (c) [(g)] diagonal components
and (d) [(h)] off-diagonal components of the optical conductivity tensor of
bulk [ML] Cr$_2$Ge$_2$Te$_6$ in the FM state with out-of-plane magnetization.
All the spectra have been convoluted with a Lorentzian of 0.2 eV to
simulate the finite electron lifetime effects.}
\end{figure}

\subsection{Optical conductivity}
We calculate the optical conductivity tensors for all the considered Cr$_2$Ge$_2$Te$_6$ 
materials with the scissor corrections. Calculated optical conductivity elements are displayed
in Fig. 6 for bulk and ML Cr$_2$Ge$_2$Te$_6$ as well as in Fig. S4 for the BL and in Fig. S5 for
the TL in the SM~\cite{SM}. Overall, the calculated optical spectra for all the systems are 
rather similar (Fig. 6, Fig. S4 and Fig. S5). This could be expected from the weak interlayer interaction
in layered vdW materials such as  Cr$_2$Ge$_2$Te$_6$.
Therefore, in what follows, we will take that of bulk and ML Cr$_2$Ge$_2$Te$_6$ as examples to 
perform detailed analysis (Fig. 6).  
First of all, the diagonal elements $\sigma_{xx}$ (for in-plane electric field polarization 
$E\perp c$) and $\sigma_{zz}$ (for out-of-plane polarization $E\parallel c$) 
of the optical conductivity for both systems are significantly different (Fig. 6),
i.e., these materials exhibit rather strong optical anisotropy.
For example, the absorptive part of $\sigma^1_{xx}$ is much larger than $\sigma^1_{zz}$ in the low energy range
of 1.0 $\sim$ 5.5 eV, while it is smaller than $\sigma^1_{zz}$ in the energy range above 6.0 eV.
This pronounced optical anisotropy could be expected from a 2D or quasi-2D material,
and also can be qualitatively explained by the Cr $d$-orbital-projected DOSs.
As mentioned above, the upper valence band ranging from -4.0 to -0.3 eV is dominated by
Cr $d$-orbitals. In particular, Figs. 5(b) and 6(b) show that in this region, the overall weight of
Cr $d_{xy,x^2-y^2}$ is much larger than that of Cr $d_{z^2}$. Given that $d_{xy,x^2-y^2}$ ($d_{z^2}$) 
states can be excited by only $ E\perp c$ ($E\parallel c$) polarized light while
Cr $d_{xz,yz}$ states can be excited by light of both polarizations, $\sigma^1_{xx}$
would obviously be larger than $\sigma^1_{zz}$ in the low energy region,
as shown in Figs. 6(a) amd 6(e). 

\begin{figure}[htb]
\begin{center}
\includegraphics[width=3.0in]{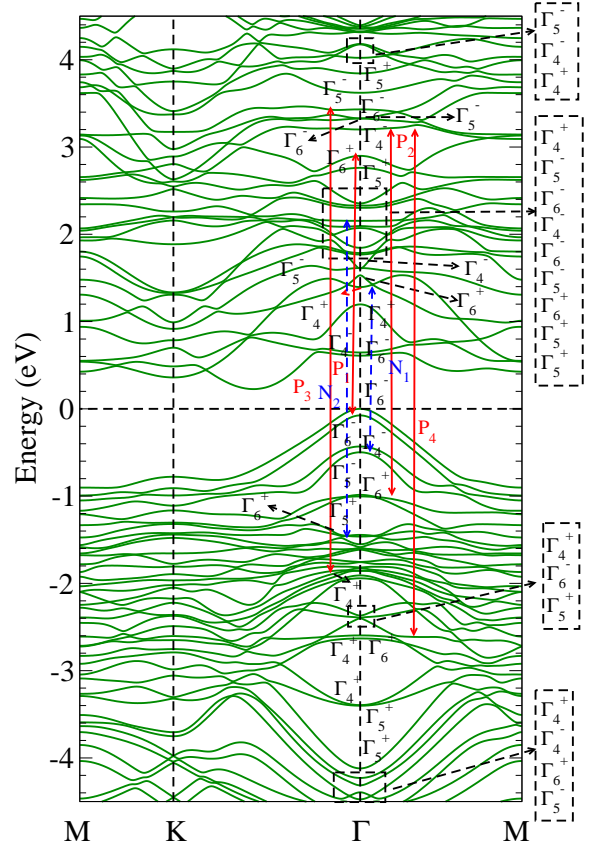}
\end{center}
\caption{Relativistic band structure of ferromagnetic monolayer Cr$_2$Ge$_2$Te$_6$ with out-of-plane magnetization.
Horizontal dashed lines denote the top of valance band. 
The symmetry of band states at the $\Gamma$-point are labelled according
to the irreducible representation of the C$_{3i}$ double point group. The principal
interband transitions and the corresponding peaks in the $\sigma_{xy}$ in
Fig. 6 (h) indicated by green arrows.}
\end{figure}

For both bulk and ML Cr$_2$Ge$_2$Te$_6$, the $\sigma_{xx}^1$ increases steeply from
the absorption edge ($\sim$1.0 eV) to $\sim$3.0 eV, and then further increases up to $\sim$4.9 eV
with a much smaller slope. Beyond $\sim$4.9 eV, the $\sigma_{xx}^1$ decreases
monotonically with the energy (see Fig. 6). The $\sigma_{zz}^1$ also increases steadily
from the absorption edge to $\sim$6.5 eV with a smaller slope, however, and then decreases
as the energy further increases (Fig. 6). 
The $\sigma_{xx}^2$ and $\sigma_{zz}^2$ spectra from bulk and ML Cr$_2$Ge$_2$Te$_6$ 
share several common characteristics: (a) a broad valley centered at $\sim$2.5 eV ($\sim$4.0 eV) 
in the $\sigma_{xx}^2$ ($\sigma_{zz}^2$), (b) a sign change 
in the $\sigma_{xx}^2$ ($\sigma_{zz}^2$) occurring at $\sim$4.5 eV ($\sim$6.5 eV), 
and (c) a plateau in the $\sigma_{xx}^2$ above 6.0 eV [see Figs. 6(c) and 6(g)]. 

Figure 6 shows that both real ($\sigma_{xy}^1$) and imaginary ($\sigma_{xy}^2$) parts of the off-diagonal
element of the optical conductivity of bulk and ML Cr$_2$Ge$_2$Te$_6$ are also rather similar. 
For example, all these spectra exhibit pronounced oscillatory peaks 
although the peak amplitude decreases with the energy. Notably, both $\sigma_{xy}^1$ and $\sigma_{xy}^2$
show a large positive peak at $\sim$2.8 eV and $\sim$3.0 eV, respectively (Fig. 6).
They also have a pronounced negative peak at $\sim$3.5 eV and $\sim$4.0 eV, respectively.

As Eqs. (4), (5) and (11) suggested, the absorptive parts of the optical conductivity elements 
($\sigma_{xx}^1$, $\sigma_{zz}^1$, $\sigma_{xy}^2$, $\sigma_{\pm}^1$) 
are directly related to the dipole-allowed
interband transitions. This would allow us to understand the origins of the main peaks
in the $\sigma_{xx}^1$, $\sigma_{zz}^1$ and $\sigma_{xy}^2$ spectra by determining the symmetries
of the calculated band states and also the dipole selection rules. 
As discussed above, the features in the $\sigma_{xx}^1$, $\sigma_{zz}^1$ and $\sigma_{xy}^2$ spectra
for all the considered systems are rather similar. Therefore, as an example, here we perform 
a symmetry analysis for ML Cr$_2$Ge$_2$Te$_6$ only (see the SM~\cite{SM}). 
The found symmetries of the band states at the $\Gamma$-point of the scalar-relativistic and relativistic
band structures of ML Cr$_2$Ge$_2$Te$_6$ are displayed in Fig. S6 and Fig. 7, respectively.
Using the dipole selection rules (Table S3), we could assign the main peaks in 
the $\sigma_{xx}^1$ and $\sigma_{zz}^1$ spectra [Fig. 6(e)] to the interband transitions 
at the $\Gamma$-point displayed in Fig. S6. For example, we could relate the A$_1$ peak at 3.4 eV
in $\sigma_{xx}^1$ [Fig. 6(e)] to the interband transition from the $\Gamma_3^-$ state
at the top of the spin-down valence band to the conduction band state $\Gamma_3^+$ 
at $\sim$3.1 eV [Fig. S6(b)]. Of course, in addition to this, there may be contributions 
from different interband transitions at other $k$-points. Note that without the SOC, the $\Gamma_3^+$
and $\Gamma_3^-$ band states are doubly degenerate (Fig. S6), and the absorption rates for left- 
and right-handed polarized lights are the same. When the SOC is included, these band states
split (Fig. 7), and this results in magnetic circular dichroism. 
Therefore, we could assign the main peaks in the $\sigma_{xy}^2$ to the principal interband 
transitions at the $\Gamma$-point only in the relativistic band structure, e. g., displayed in Fig. 7.
In particular, we could attribute the pronounced peak P$_1$ at $\sim$3.0 eV in $\sigma_{xy}^2$ [Fig. 6(h)]
to the interband transition from the $\Gamma_4^-$ and $\Gamma_6^-$ states at the top of the valence
band to the conduction band state $\Gamma_6^+$ at $\sim$2.8 eV (Fig. 7).

\begin{figure}[htb]
\begin{center}
\includegraphics[width=7.5cm]{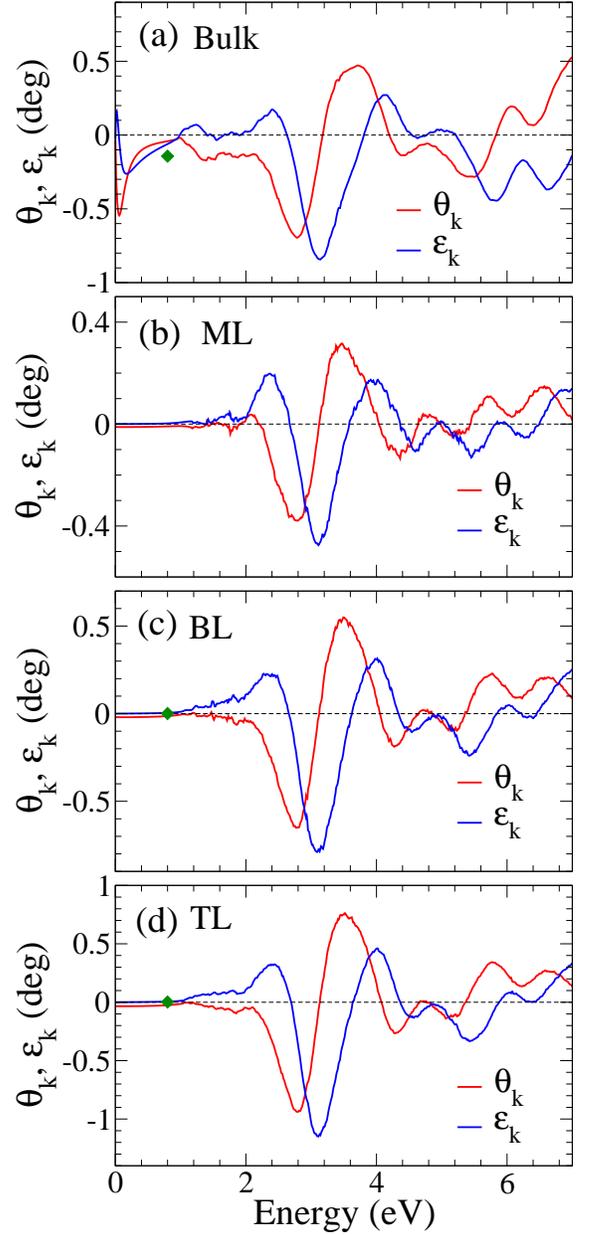}
\end{center}
\caption{Kerr rotation ($\theta_K$) amd ellipticity ($\varepsilon_K$) spectra for (a) bulk, (b) ML, 
(c) BL, and (d) TL Cr$_2$Ge$_2$Te$_6$ in the FM state with out-of-plane magnetization. 
Green diamonds denote the $\theta_K$ value from the recent experiments~\cite{gong2017}}	
\end{figure}

\subsection{Magneto-optical Kerr and Faraday Effects}
Here we consider the polar Kerr and Faraday effects for all the systems considered, and calculate 
their complex Kerr and Faraday rotation angles as a function of photon energy, as plotted 
in Figs. 8 and 9, respectively. Figure 8 shows that for the three multilayers, the spectra 
are negligibly small below $\sim$1.2 eV but become large and oscillatory in the incident 
photon energy range from 1.2 eV to 7.0 eV. In particular, the Kerr rotation angle
in all the multilayers, is remarkably large in the vicinity of 2.8 eV, reaching up to 0.4{\degree}
for the ML, 0.6{\degree} for the BL and 1.0{\degree} for the TL. The shapes of the Kerr rotation
spectra for all the multilayers are similar, indicating the weakness of the vdW
interlayer interaction. The Kerr rotation spectrum for the bulk is similar to that of
the multilayers, except it is not negligible below 1.2 eV where bulk Cr$_2$Ge$_2$Te$_6$ 
also shows significant Kerr rotations [see Fig. 8(a)], which will be explained below.

A comparison of Fig. 8 with Fig. 6 would reveal that the Kerr rotation ($\theta_K$) and Kerr ellipticity
($\varepsilon_K$) spectra in all the systems resemble, respectively, the corresponding
real part ($\sigma^1_{xy}$) and imaginary part ($\sigma^2_{xy}$) of the off-diagonal conductivity element.
This is not surprising because the Kerr effect and the off-diagonal conductivity element
are connected via Eqs. (8) and (9). Indeed, Eqs. (8) and (9) indicate that the complex Kerr rotation
angle would be linearly related to the $\sigma_{xy}$ if the longitudinary conductivity ($\sigma_{xx}$)
of the bulk and substrate is more or less constant. For the Cr$_2$Ge$_2$Te$_6$ multilayers, the latter is
true because here we assume that the substrate is SiO$_2$ with dielectric constant $\varepsilon_s = 3.9$.
For bulk Cr$_2$Ge$_2$Te$_6$, the Kerr rotation could become large if the
$\sigma_{xx}$, which is in the denominator of Eq. (8), becomes very small.
This explains that the Kerr rotation of the bulk is still visible below 1.2 eV (Fig. 8).
  
Let us now compare the calculated Kerr rotation angles in the Cr$_2$Ge$_2$Te$_6$ materials 
with that found in several well-known MO materials.
Ferromagnetic 3$d$ transition metals and their alloys form an important class 
of metallic MO materials.~\cite{antonov2004}
Kerr rotation angles of these metals seldom exceed 0.5{\degree}
except a few of them such as heavy element Pt intermetallics 
FePt, Co$_2$Pt ~\cite{guo1996} and PtMnSb~\cite{van1983ptmnsb}
where the strong SOC in the Pt atoms plays an important role.~\cite{guo1996}.
Manganese pnictides also have excellent MO properties. In particar, MnBi films possess a large
Kerr rotation angle of 2.3{\degree} at 1.84 eV in low temperatures.~\cite{di1996optical,ravindran1999}
Remarkably, calculated Kerr rotation angles of bulk and few-layer Cr$_2$Ge$_2$Te$_6$
(Fig. 8) are comparable to that of these traditional excellent MO materials.
Furthermore, they are generally more than 10 times larger than
that of FM 3$d$ transition metal MLs deposited on metallic substrates.
For example, BL Fe epitaxially grown on Au (001) surface exhibits 
largest Kerr rotation angle of only $\sim$0.025{\degree} at $\sim$2.75 eV.~\cite{geerts1994}  
In this context, the Kerr rotation angles of atomically thin Cr$_2$Ge$_2$Te$_6$ films 
could be regarded to be gigantic.

Among famous MO semiconductors, Y$_3$Fe$_5$O$_{12}$ exhibits a significant Kerr rotation 
of 0.23{\degree} at 2.95 eV.~\cite{tomita2006} Recently, dilute magnetic semiconductors were 
reported to show significant Kerr rotation angles of $\sim$0.4{\degree}
in the vicinity of 1.80 eV.~\cite{lang2005}
Remarkably, Feng {\it et. al.} recently studied theoretically the MO properties of hole-doped Group-IIIA
metal-monochalcogenide MLs and predicted that many of these FM MLs would 
exhibit significant Kerr rotations of $\sim$0.3{\degree} at optimal hole concentrations.~\cite{feng2016}
On the whole, the Kerr rotation angles predicted for bulk and
few-layer Cr$_2$Ge$_2$Te$_6$ here are comparable to these important MO semiconductors.
Therefore, because of their excellent MO properties, Cr$_2$Ge$_2$Te$_6$ materials could 
find promising applications for, e.g., MO sensors and high density MO data-storage devices.
Moreover, MOKE in atomically thin films as well as bulk Cr$_2$Ge$_2$Te$_6$ at low temperatures has been
measured using highly sensitive Sagnac interferometer with light wavelength of 1550 nm 
(photon energy of 0.8 eV).~\cite{gong2017}
The measured $\theta_K$ value of $\sim$0.14{\degree} of the bulk is in the same order of magnitude with our theoretical
prediction [see Fig. 8(a)]. The $\theta_K$ values for the thin films are, however, much smaller,
ranging from 0.0007{\degree} in bilayer to 0.002{\degree} in trilayer.~\cite{gong2017}  
Such small measured $\theta_K$ values could be attributed to the fact that the energy of light beam used
falls almost within the bandgap where MOKE is negligibly small (Fig. 8).  
Indeed, the $\theta_K$ value of $\sim$0.28{\degree} of the CrI$_3$ ML measured using 
a 633-nm HeNe laser (photon energy of 1.96 eV)~\cite{huang2017} is orders of magnitude larger than
the Cr$_2$Ge$_2$Te$_6$ thin films reported in ~\cite{gong2017}.

\begin{figure}[htb]
\begin{center}
\includegraphics[width=7.5cm]{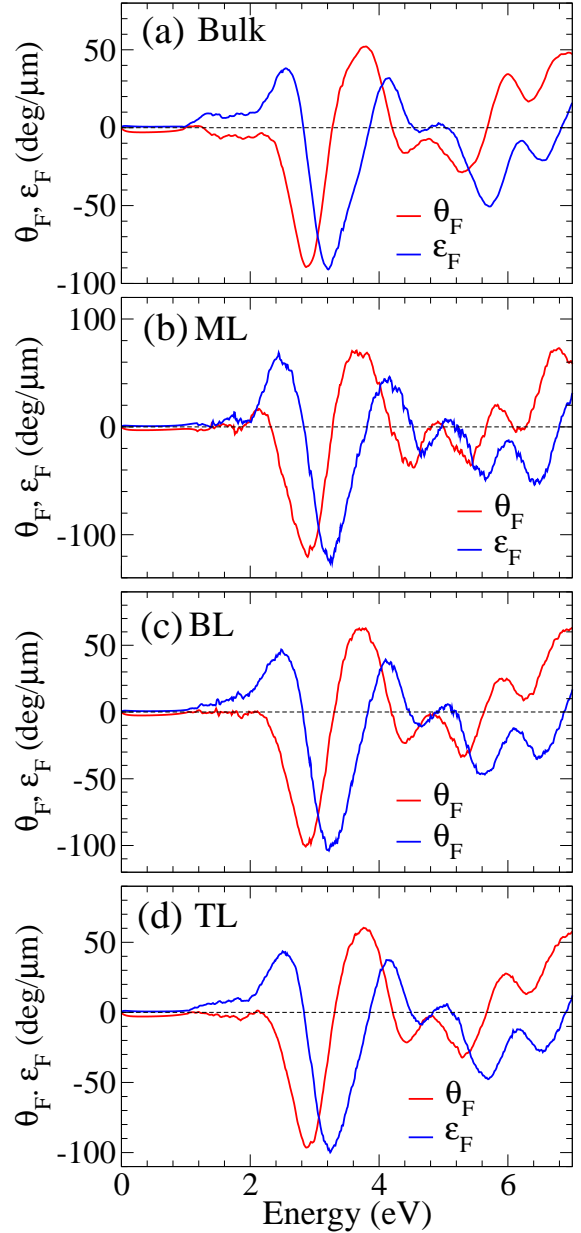}
\end{center}
\caption{Faraday rotation ($\theta_F$) amd ellipticity ($\varepsilon_F$) spectra for (a) ML, 
(b) BL and (c) TL Cr$_2$Ge$_2$Te$_6$ in the FM state with out-of-plane magnetization.}	
\end{figure}

Complex Faraday rotation angles for all the considered structures are displayed
in Fig. 9. The Faraday rotation spectra are similar to the corresponding Kerr rotation spectra (Fig. 8)
as well as the $\sigma_{xy}$ (see Fig. 6). Figure 6 shows that the $\sigma_{xx}$ is generally much larger than
the $\sigma_{xy}$. Thus, $n_{\pm }=[1+{\frac{4\pi i}{\omega}}(\sigma _{xx}\pm i \sigma _{xy})]^{1/2}$
$\approx [1+{\frac{4\pi i}{\omega}}\sigma _{xx}]^{1/2} \mp {\frac{2\pi}{\omega}}(\sigma _{xy}/\sqrt{1+\frac{4\pi i}{\omega}\sigma _{xx}})$. 
Consequently, Eq. (10) can be approximately written as 
$\theta _{F}+i\epsilon _{F}\approx -\frac{2\pi d}{c}(\sigma _{xy}/\sqrt{1+\frac{4\pi i}{\omega}\sigma _{xx}})$
or 
$\theta _{F}+i\epsilon _{F}\approx -\frac{2\pi d}{c}\frac{\sigma _{xy}}{\sqrt{\varepsilon_{xx}}}$
where $\varepsilon_{xx}$ is the dielectric function.
This explains why the complex Faraday rotation follows closely $\sigma _{xy}$ (Figs. 6 and 9).

Remarkably, calculated maximum Faraday rotation angles are as large as $\sim 120$ deg/$\mu$m in 
2D Cr$_2$Ge$_2$Te$_6$ (see Fig. 9). 
In the visible frequency range (below 4.0 eV), they are more than ten times larger than 
the predicted Faraday rotations in Group-IIIA
metal-monochalcogenide monolayers at optimal hole dopings.~\cite{feng2016}
They are even comparable to that of prominent bulk MO metals such as manganese pnictides. 
In particular, among manganese pnictides, MnBi films possess the largest Faraday rotations of $\sim 80$ deg/$\mu$m 
at 1.77 eV at low temperatures.~\cite{di1996optical,ravindran1999}
Even the famous MO semiconductor Y$_3$Fe$_5$O$_{12}$ exhibits a Faraday rotation 
only as large as $0.19$ deg/$\mu$m at 2.07 eV.\cite{boudiar2004} 
However, substituting Y with Bi could substantially enhance Faraday rotations up to 
$\sim35.0$ deg/$\mu$m at 2.76 eV in Bi$_3$Fe$_5$O$_{12}$~\cite{vertruyen2008curie}. 
Clearly, the Faraday rotation angles reported here for few-layer multilayer Cr$_2$Ge$_2$Te$_6$ 
are comparable or even superior to many well-known MO materials. This suggests that one could also
exploit the large Faraday effect to probe the long range magnetic orders in these
quasi-2D magnetic materials. Furthermore, benefited from the excellent MO properties, 
Cr$_2$Ge$_2$Te$_6$ materials could find valuable applications for magnetic-optical devices.

\section{CONCLUSION}
We have investigated magnetism as well as electronic, optical and
magneto-optical properties of atomically thin FM films recently exfoliated from 
bulk FM semiconductor Cr$_2$Ge$_2$Te$_6$ by performing systematic GGA+U calculations.
In particular, we focus on two relativity-induced properties of these 2D materials, 
namely, magnetic anisotropy energy and MO effects.
Firstly, we find that calculated MAEs of these materials are large, being in the order of $\sim$0.1 meV/Cr. 
Interestingly, the out-of-plane anisotropy is found in all the considered systems 
except the ML. In contrast, in the ML an in-plane magnetization is preferred simply 
because the D-MAE is larger than the C-MAE. Crucially, this would explain why long-range
FM order was recently observed in all the few-layer Cr$_2$Ge$_2$Te$_6$ except the ML.
This is because the out-of-plane anisotropy would open a spin-wave gap and thus suppress magnetic fluctuations 
so that long-range FM order could be stabilized at finite temperature.
Secondly, large Kerr rotations up to $\sim$1.0{\degree}
in these FM materials are found in the visible frequency range, and they are comparable
to that observed in famous MO materials such as FM metal PtMnSb and
semiconductor Y$_3$Fe$_5$O$_{12}$. Moreover, they are two-order of magnitude larger than
that of 3$d$ transition metal MLs deposited on Au surfaces, and thus can be called gigantic.
Thirdly, calculated maximum Faraday rotation angles in these 2D materials are also large, being up
to $\sim$120 deg/$\mu$m, and are comparable to the best-known MO semiconductor Bi$_3$Fe$_5$O$_{12}$. 
These findings thus suggest that with large MO effects
plus significant MAE, atomically thin films of Cr$_2$Ge$_2$Te$_6$ might 
find valuable applications in 2D magnetic, magneto-electric and MO device
such as high-density data-storage and nanomagnetic sensors.
Fourthly, calculated Kerr rotation angles at 1550 nm wavelength
are in reasonable agreement with recent MO Kerr effect experiments.
The FM transition temperatures estimated using the calculated exchange coupling parameters
within the mean-field approximation plus the spin-wave gap correction agree quite well with
the measured transition temperatures. Finally, the calculated C-MAE and MO properties of these 2D
materials are analyzed in terms of their electronic band structures.

\section*{ACKNOWLEDGEMENTS}
Y. F. thanks Department of Physics and Center for Theoretical Physics, National Taiwan University 
for its hospitality during her three months visit there when parts of this work were carried out.
Work at Xiamen University is supported by the National Key Research Program of China (Grant No. 2016YFA0202601),
the National Natural Science Foundation of China (No. 11574257), and the Fundamental Research Funds for the
Central Universities (No. 20720180020).
G. Y. G. acknowledges the support by the Ministry of Science and Technology, the Academia Sinica, 
the National Center for Theoretical Sciences and the Kenda Foundation of Taiwan.



\end{document}